\documentclass{aa}
\input psfig.sty
\usepackage{graphicx}
\usepackage{txfonts}
\begin{document}

   \title{Improvement of the CORS method for Cepheids radii determination \\ based on Str\"omgren photometry }

   \subtitle{}

   \author{A. Ruoppo
          \inst{1,2}
          \and
          V.Ripepi
          \inst{1}
          \and
          M.Marconi
          \inst{1}
          \and
          G.Russo
         \inst{2}
          }

    \offprints{Vincenzo Ripepi}

   \institute{Osservatorio Astronomico di Capodimonte, via Moiariello 16, 80131 Napoli, Italy\\
              \email{ruoppo@na.astro.it, ripepi@na.astro.it, marcella@na.astro.it}
         \and
             Dipartimento di scienze fisiche universit\`a di Napoli Federico II, Complesso Monte S.Angelo, 80126 Napoli, Italy\\
             \email{ale@na.infn.it}, \email{guirusso@unina.it}
             }

   \date{ }

   \abstract{In this paper we present a modified version of the CORS 
method based on a new calibration of the Surface Brightness 
function in the Str\"omgren photometric system. The method has 
been tested by means of synthetic light and radial velocity curves 
derived from nonlinear pulsation models. Detailed simulations have been 
performed to take into account the quality of real observed curves 
as well as possible shifts between photometric and radial velocity data.
The method has been then applied to a sample of Galactic Cepheids 
with Str\"omgren photometry and radial velocity data to derive the 
radii and a new PR relation. As a result we find 
$\log R = (1.19 \pm 0.09) + (0.74 \pm 0.11) \log P$ (r.m.s=0.07).
The comparison between our result and previous estimates 
 in the literature is satisfactory.
Better results are expected from the adoption of improved 
model atmosphere grids.
      \keywords{stars: distance --
                stars: fundamental parameters --
                stars: variables: Cepheids
               }
   }

   \maketitle
%

\section{Introduction}

Classical Cepheids are the cornerstone of the extragalactic distance scale. 
Thanks to their characteristic Period-Luminosity (PL) and 
Period-Luminosity-Color (PLC) relations they are traditionally 
used to derive the distances to Local Group galaxies, 
and (with the advent of space observations) to external galaxies distant up 
to about 25 Mpc (targets of a Hubble Space Telescope Key Project, see Freedman et al. 1997, 2001). As primary indicators they are used to calibrate 
a number of secondary distance indicators (see e.g. Freedman et al. 2001) 
reaching the region of the so called {\it Hubble flow} where 
the Hubble law can be applied and an estimate of the Hubble 
constant can be derived.\\ 
Moreover, the comparison between Cepheid physical parameters 
(stellar mass, luminosity, chemical composition) 
based on evolutionary and pulsation models supplies
the unique opportunity to pin point the occurrence 
of deceptive systematic errors (Bono et al. 2001a; Moskalik 2000) on the 
Cepheid distance scale.\\
In particular radius determinations are important to constrain 
both the intrinsic luminosity, through the application of the 
Stefan-Boltzmann law, provided that an effective temperature calibration 
is available, and the stellar mass, by adopting a Period-Mass-Radius 
relation (e.g. Bono et al. 2001).\\
Many investigations have been devoted during the last decade to the 
derivation of accurate Period-Radius (PR) relations for Classical Cepheids 
both from the empirical (see e.g. Laney \& Stobie 1995; Gieren, 
Fouqu\'e, \& Gomez 1998; Ripepi et al. 1997) and the 
theoretical (Bono, Caputo, Marconi 1998; Marconi et al. 2003) point of view.\\
Empirical Cepheid radii are generally derived either by means of 
the Baade Wesselink (BW) method (Moffet \& Barnes 1987, Ripepi et al. 1997;
Gieren et al. 1998, just to list a few examples) both in the classical 
form and in subsequent modified versions, or with interferometric coupled with trigonometric parallaxes 
techniques (Nordgren, Armstrong \& German 2000, Lane, Creech-Eakman \& Nordgren 2002).\\
The latter method is more direct and less model dependent but up to now 
it has been applied only to a limited number of stars. 
On the other hand, the different versions of the BW technique can be 
applied to relatively large Cepheid samples but require 
both accurate photometric and radial velocity data.\\
A powerful modification of the BW technique is the so called CORS 
method (Caccin et al. 1981), which has the advantage of taking into 
account the whole light curve rather than selecting phase points 
at the same color (as in the classical BW implementation), but relies on 
the adoption of an accurate Surface Brightness (SB) calibration.\\
Originally Sollazzo et al., (1981) adopted the empirical SB photometric 
calibration in the Walraven system provided by Pel (1978). More recently
Ripepi et al. (1997) modified the method, by adopting the empirical calibration 
of the reduced surface 
brightness $F_V$  as a function of $(V-R)$ provided by 
Barnes, Evans $\&$ Parson  (1976).
This modified version of the CORS method was tested, for different colors 
selections, through the 
application to synthetic light and radial velocity curves based 
on nonlinear convective pulsation models (Ripepi et al. 2000).\\ 
The recent release of new Cepheid data in the Str\"omgren 
photometric system (Arellano-Ferro et al. 1998), and the known sensitivity 
of intermediate band colors to stellar physical parameters 
(e.g. gravity and  effective temperature)  suggested us to 
investigate the possibility of extending the CORS method to the 
Str\"omgren filters.\\
To this purpose we have derived, in this system, a SB calibration based 
on model atmosphere tabulations. In this paper we present a modified 
version of the CORS method based on this new calibration and the 
application to a sample of Galactic Cepheids.\\
The organization of the paper is the following: in Sect. 2 we summarize 
the assumptions and the philosophy of the traditional CORS method; 
in Sect. 3 we introduce the modified CORS method based on the new 
SB calibration; in Sect. 4 we test the new method by means of pulsation 
models; in Sect. 5 we apply the method to a sample of Galactic Cepheids 
and  present the comparison with the literature; in Sect. 6   our 
final results concerning  the PR relation for Galactic Cepheids are shown
and the theoretical fit of observed light and radial velocity variations
for Cepheid Y Oph is used as an additional check. Some final remarks 
close the paper.


\section{ The original CORS method}
In this section we briefly outline the assumptions and the main features of the 
CORS method in order to 
better understand what follows.\\  
The CORS method (Caccin et al., 1981) starts from the definition of the surface brightness:
\begin{equation}
\label{eq:bs}
S_V = V + 5 \cdot \log \alpha 
\end{equation}
where $\alpha$ is the angular diameter of the star.
For a variable star,  Eq.(\ref{eq:bs}) is valid for the whole 
pulsational cycle, so that 
differentiating it with respect to the phase ($\phi$), 
multiplying by a color index ($C_{ij}$) and integrating 
over the whole cycle, one obtains:
\begin{eqnarray} 
\label{eq:eqcors}
\lefteqn{q \int_{0}^{1}ln \ \Big\{ R_0 - k \ P \int_{\phi_0}^{\phi} v(\phi ') \ d \phi ' \Big\} \ C_{ij}' d \phi} \nonumber \\
& & - B + \Delta B =0
\end{eqnarray}
where $q=\frac{5}{\ln 10}$
\begin{equation}
\label{eq:B}
B = \int_{0}^{1} C_{ij}(\phi) \ m_V '(\phi) \ d\phi
\end{equation}
\begin{equation}
\label{eq:deltaB}
\Delta B = \int_{0}^{1} C_{ij}(\phi) \ S_V '(\phi) \ d\phi
\end{equation}
where $P$ is the period, $v$ the radial velocity and $k$ is the 
radial velocity projection factor which relates radial to pulsation
velocity 
($R'(\phi)= -k \cdot P \cdot v(\phi)$). The typical value for $k$
is 1.36 (see discussion 
in Ripepi et al. 1997).\\ 
From  Eq.(\ref{eq:eqcors}) we can evaluate the radius $R_0$
at an arbitrary phase $\phi_0$ (usually 
taken at the minimum of the radial velocity curve), whereas 
the mean radius is obtained by integrating twice the radial velocity curve. \\
The $B$ term  can be easily calculated from observed light 
and color curves. On the contrary, the $\Delta B$ term, which   
includes the Surface Brightness, is not directly observable. 
By neglecting it in Eq. (\ref{eq:eqcors}), 
we obtain the  pure Baade-Wesselink method (see Caccin et al. 1981).
However,  Sollazzo et al. (1981) and Ripepi et al. (1997,  
 hereinafter RBMR)
 demonstrated that the inclusion of $\Delta B$ improves the accuracy 
of radius estimates, provided that $S_V$ is evaluated 
at each pulsation phase. \par


\section{ An Improvement of the CORS method based on the 
Str\"omgren photometry }

\subsection{Evaluation of the $\Delta$B term}

As outlined in the previous section, the inclusion of $\Delta B$ 
improves the accuracy of radius estimates. In this section we present a new
 good approximation for this term.

As discussed by Onnembo et. (1985), if the quasi-static approximation 
(QSA) \footnote{We recall 
that a sufficient condition for the validity of the QSA is that 
the atmosphere of the pulsating star can be described, at any time, by a classical
hydrostatic, plane parallel model in radiative/convective 
equilibrium and local-thermodynamic-equilibrium (LTE), identified by the effective temperature 
$T_{eff}$  and by the effective gravity 
$g_{eff}= \frac{G M}{R^2}+ \frac{d^2 R}{dt^2}$} is assumed  for 
Cepheid atmospheres, any photometric quantity can be expressed 
as a function of effective temperature and gravity ($T_{eff}, g_{eff}$); 
Then we can write:
\begin{equation}
\label{eq:sb}
S_V = S_V (T_{eff}, g_{eff})
\end{equation}
and
\begin{eqnarray}
\label{eq:color}
c_{ij} = c_{ij} (T_{eff}, g_{eff})\\
c_{kl} = c_{kl} (T_{eff}, g_{eff})
\nonumber
\end{eqnarray}
where $S_V$ is the surface brightness in the visual band, 
and $c_{ij} = (m_i -m_j)$, $c_{kl} = (m_k -m_l)$ are two arbitrary colors.\\
If the last two equations are invertible, i.e 
the Jacobian $J (T_{eff}, g_{eff}| C_{ij}, C_{hl}) \ne 0$, 
then we can invert Eqs. (\ref{eq:color}), obtaining:
\begin{eqnarray} 
\label{eq:tg}    
 T_{eff} = T_{eff}(c_{ij}, c_{kl})\\
 g_{eff} = g_{eff}(c_{ij}, c_{kl})
\nonumber 
\end{eqnarray}
and hence 
\begin{equation} 
\label{eq:sv}
S_V = S_V (c_{ij}, c_{kl})
\end{equation}
We point out that Eqs. (\ref{eq:color}) do not admit a general 
solution over the whole parameter space, given that the same colors 
can be obtained for different pairs of $T_{eff}$ and $g_{eff}$. 
This notwithstanding it is possible to find a local solution, 
that is a solution valid only in the parameter space defined
by Cepheids. \par

A potentially advantageous choice for the colors $c_{ij}$, $c_{kl}$ could be  
represented by the Str\"omgren reddening free indexes $[m_1]$ and $[c_1]$ 
defined as follows (Crawford \& Mandwewala 1976):

\begin{equation}
\label{eq:m1q}
[m_1] = m_1 + 0.33 (b-y)
\end{equation}
\begin{equation}
\label{eq:c1q}
[c_1] = c_1 - 0.16 (b-y)
\end{equation} 
where the coefficients $0.33$ and $-0.16$ are suitable for F supergiant stars (Gray 1991).\\

\begin{figure}
   \centering
   \includegraphics[width=8cm]{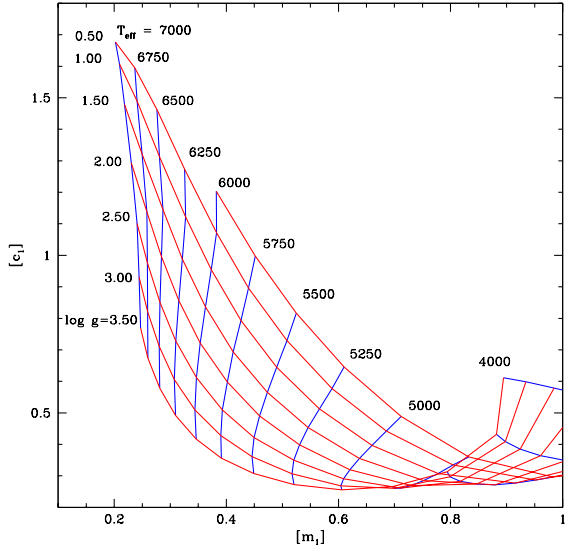}
      \caption{A grid of lines at constant $T_{eff}$ and constant $\log g $ in the theoretical $[m_1]$, $[c_1]$ plane.\label{griglia1}}
         
   \end{figure}

In this context, an interesting possibility to derive Eqs. 
(\ref{eq:tg}), consists in using grids of theoretical colors, 
calculated by means of model atmospheres, 
to obtain $T_{eff}$ and $g_{eff}$ as a function of $[m_1]$ and $[c_1]$.\\
To verify this possibility, we have adopted the grids of 
theoretical colors by Castelli, Gratton \& Kurucz 
(1997a,b, hereinafter C97a,b). 
The theoretical grid (constant $T_{eff}$ and $\log g$) in the 
 $[m_1]$, $[c_1]$ plane is shown in Fig. \ref{griglia1}. This figure 
suggests that for
$0.50\le\log g\le 3.50$ and $5000 K \le T_{eff} \le 7000 K $ there 
is a one-to-one correspondence between a point in the $[m_1]$, $[c_1]$ plane 
and the corresponding $T_{eff}$ and $\log g$ values. 
Thus, in principle, in this color range it is possible to 
invert Eqs. (\ref{eq:color}) and, in turn, to derive an expression for 
Eq.(\ref{eq:sv}).\\ 
 
\begin{figure}
   \centering
   \includegraphics[width=8cm]{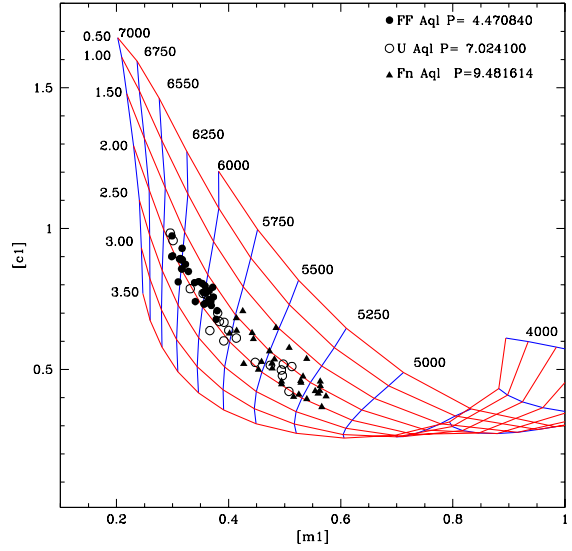}
      \caption{Theoretical grids in the $[m_1],[c_1]$ plane, compared with
 the empirical loops of three stars with  period shorter than 10 d. \label{griglia2}}
   \end{figure}
\begin{figure}
   \centering
   \includegraphics[width=8cm]{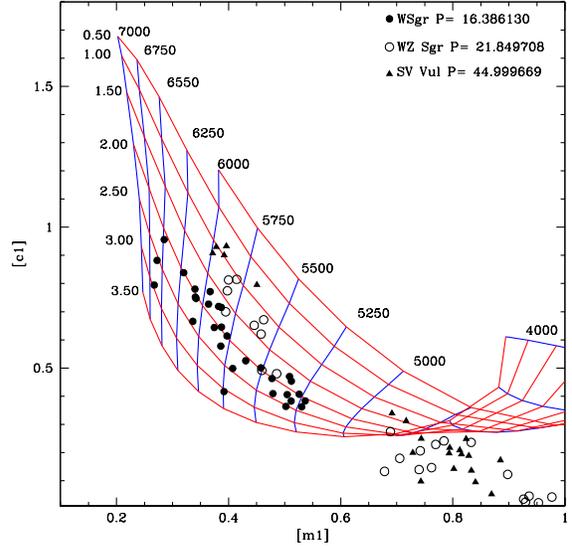}
      \caption{The same as in Fig. 2 but for three Cepheids with period longer than 15 d.\label{griglia3}}
         
   \end{figure}

Before proceeding, it is important to verify if the location of 
theoretical grids in the $[m_1],[c_1]$ plane is consistent with the one 
occupied by  real Cepheid data. 
To this aim, we have overplotted on Fig.~\ref{griglia1} 
the color-color $[m_1],[c_1]$ loop for all the stars in our sample 
(see section 5.1). Figure \ref{griglia2} shows the resulting comparison
 for three stars (FF Aql, FN Aql, U Aql) of our sample, 
characterized by short and intermediate periods. The figure shows that  
the color-color loops for the selected stars are completely included 
in the theoretical grids. The same test has been 
performed for the three stars W Sgr, WZ Sgr, SV Vul, 
which have longer periods. As showed by Fig.~\ref{griglia3}, in this case 
 the color-color loops for the selected Cepheids 
lies outside (at low gravity and low effective temperature) of the region covered by theoretical grids. This unexpected 
result has to be taken into account when applying our method 
to such long period stars.
An explanation for this model limitation is beyond the scopes of present
 paper but it is an important issues worth to be addressed in a future work.\\
 
 
\subsubsection{Derivation of $S_V$}

As shown in the previous section, the first step in the construction of a new, based on Str\"omgren photometry, version of the CORS method, is the formulation of  $S_V$ in the form of 
Eq.(\ref{eq:sv}). To this aim, we have first to find relations in the form of 
Eqs.(\ref{eq:tg}) starting from theoretical grids. 
This has been achieved by means of a $4^{\textit{th}}$ degree polynomial fit 
to the theoretical grid (least square fit). The results of the 
polynomial approximations are reported in  Appendix A, 
and showed in  Figs. \ref{fig2} and \ref{fig3} for $\log T_{eff}$ and 
$\log g_{eff}$ respectively. The rms of the fits are:0.0018 dex for $\log T_{eff}$  
and 0.1 for $\log g_{eff}$.\\

On the basis of the calculated relations we are now in the position to 
estimate the surface brightness from the 
expression: 

\begin{equation}
\label{eq:sf}
S_V = const. - 10 \log T_{eff} - BC(T_{eff},g_{eff})
\end{equation}
where BC is the Bolometric Correction, derived as a function of effective temperature and gravity ($BC=BC(T_{eff},g_{eff})$). This function can be easily obtained
using again a $4^{\textit{th}}$ degree polynomial fit 
to the theoretical grids. The resulting equation is also reported in  
Appendix A and shown in Fig.\ref{supBC}. Note that the rms is only 0.003 mag.

The procedure outlined above allowed us to achieve our goal, i.e. to derive 
an analytic, although approximated, expression of $S_V = S_V ([m_1], [c_1])$, 
and, in turn, to calculate the $\Delta B$ term, which allows us to apply the 
CORS 
method, in its more general formulation, for data in the Str\"omgren photometric system.\\

   \begin{figure}
   \centering
      \caption{ $\log T_{eff}$ as a function of $[m_1], [c_1]$ obtained by a $4 ^{\textit{th}}$ degree polynomial fit of theoretical data (crosses)  by C97a,b. \label{fig2}}
        
   \end{figure}

   \begin{figure}
   \centering
      \caption{ The same as in Fig. 4 but for $\log g$.\label{fig3}}
         
   \end{figure}

\begin{figure}[htbp!]
\begin{center}
\caption{\footnotesize $BC$ as a function of $\log T_e$, $\log g$ obtained by a $4 ^{\textit{th}}$ degree polynomial fit of theoretical data (crosses) by C97a,b. \label{supBC}}
\end{center}
\end{figure}
\section{Test of the revised CORS method}
To verify the accuracy of the new approximation, 
we have applied the CORS method (with the new $S_V$ calibration) 
to  synthetic light, color and radial velocity curves predicted 
by Cepheid full amplitude, nonlinear, convective models. 
The advantages of testing the method by means of pulsation model curves, rather than 
empirical data, are the following:
\begin{enumerate}
 \item for pulsation models, bolometric light curves are transformed into
 the observational bands using the same model atmosphere tabulations
 (C97a,b) we have used to derive the surface
 brightness calibration;
\item pulsation models provide a set of equally well sampled curves covering  
wide period and effective temperature ranges;
\item  for pulsation models the radius and intrinsic luminosity information 
are available.
\end{enumerate}

In the following we describe in detail how the models have been used 
to test the method.

\subsection{The synthetic curves}

To test both the accuracy and the consistency of the new CORS 
method we have adopted the pulsation observables predicted by hydrodynamical 
models of classical Cepheids. A detailed discussion on the physical 
assumptions adopted to calculate these models can be found in 
Bono, Marconi \& Stellingwerf (1999, BMS99) and Bono, Castellani, 
Marconi (2002). 
Among the different sequences of nonlinear models computed by BMS99 we have 
selected canonical models\footnote{Canonical models are the ones 
 constructed by adopting a mass-luminosity relation 
based on evolutionary computations which neglect convective 
core overshooting during hydrogen burning phases (Castellani, Chieffi \& Straniero 1992).} 
at solar chemical composition ($Y=0.28, Y=0.02$) and stellar masses 
ranging from 5 to 9 $M_{\odot}$. At fixed stellar mass, we generally 
chose three models which are located in the middle of the 
instability strip as well as close to the blue and the red edge. 
The period of selected models roughly ranges from 3.5 to 62 days. 
The input parameters ($M, T_{eff}, L$), the computed radius and the 
pulsational period are summarized in Table \ref{tabella1}.\\ 

\begin{table}[h]
\caption{Physical properties of the selected Cepheid models \label{tabella1}}
\begin{tabular}{rrccrr}
\hline
\hline\noalign{\smallskip}
model &	Mass            & Luminosity &	  $T_{eff}$ & Radius & Period \\
      & $\rm M_{\odot}$ & $\rm \log{L/L_{\odot}}$ & K & $\rm R_{\odot}$ & Days \\
\hline\noalign{\smallskip}
mod1  &  5    &3.07& 5800  &   34.3  &  3.5091     \\
mod2  &  5    &3.07& 5600  &   36.7  &  3.9412     \\
mod3  &  6.25 &3.42& 5400  &   58.6  &  7.5842     \\
mod4  &  6.25 &3.42& 5100  &   65.2  &  8.7060     \\
mod5  &  7    &3.65& 5300  &   81.0  &  12.0904    \\
mod6  &  7    &3.65& 5000  &   91.3  &  14.7582    \\
mod7  &  7    &3.65& 4800  &   97.3  &  16.8322    \\
mod8  &  9    &4.0 & 4900  &   185.7 &  46.5687    \\
mod9  &  9    &4.0 & 4700  &   204.1 &  53.9378    \\
mod10 &  9    &4.0 & 4500  &   220.6 &  62.3650    \\
\noalign{\smallskip}
\hline
\hline
\end{tabular}
\end{table}

\begin{figure}
   \centering
   \includegraphics[clip,width=6cm ]{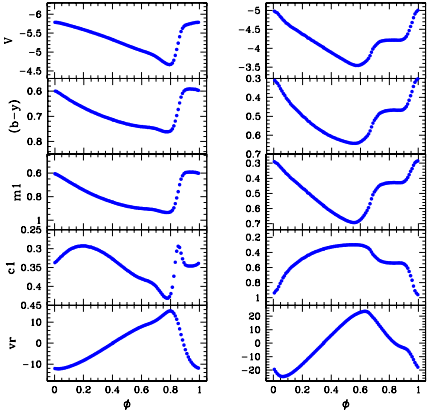}
      \caption{ Variations along a full pulsation cycle of magnitude, color and radial velocity for  models  with  $M = 9 M_{\odot}$ and $T_{eff} = 4500 K$ (left) and $M = 7 M_{\odot}$ and $T_{eff} = 5300 K$ (right). \label{curveteo}}
         
   \end{figure}

Theoretical observables have been transformed into the observational 
plane by adopting the bolometric corrections (BC) and the 
color-temperature relations by C97a,b. 
We assumed $M_{Bol} (\odot) = 4.62$ mag and adopted atmosphere models 
 computed by neglecting the core overshooting 
and for a fixed value of the microturbolence velocity 
$\xi = 2 km s^{-1}$.\\
For each pulsation model we have derived the  \textit{V} light curve,  three color curves, namely in \textit{(b-y)}, $[m_1]$ and $[c_1]$ and the radial velocity curve.  
Figure \ref{curveteo} shows these curves  for  models with  $M = 9 M_{\odot}$, $T_{eff} = 4500 K$, and $M = 7 M_{\odot}$,  $T_{eff} = 5300 K$.


\subsection{Application to theoretical models}

As a first test, we have applied the method 
to ``perfect'' light, color and radial velocity curves, i.e. 
the curves directly obtained from the models which, of course, do not 
show random errors. This 
test allows us to verify whether or not  our calibration of the surface brightness 
is  intrinsically correct. If this is the case we do not expect 
a large discrepancy between the calculated radii and the ``true'' 
(theoretical) ones.\\ To perform this comparison
 we have evaluated the radius of each model in the two 
different approximations:

\begin{itemize}
\item  Without the $\Delta B$ term (Baade-Wesselink approximation);
\item  With the $\Delta B$ estimated as described in section 3.
\end{itemize}

The results of this test are summarized in the first four columns of Table~\ref{tabella2}.  We note that, the application to models with long period is justified by the fact that the bolometric light curves of pulsation models have been transformed into the observational bands by using the same model atmospheres by C97a,b. Therefore, at variance with the observed long period Cepheids, the model color-color loops are always consistent with C97a,b's grids. 


\begin{table}[h]
\centering
\caption{Radii estimated with the two different approximations , 
i.e  with and without the $\Delta B$ term (respectively column 2,3),  using model predictions (column 4), and from modified synthetic curves in the case of good (column 5) and fair (column 6) data respectively. 
\label{tabella2}}
\begin{tabular}{cccccc}
\hline
\hline\noalign{\smallskip}
Models&R$_{without \Delta B}$&R$_{\Delta B}$&R$_{teo}$&R$^1_{\Delta B}$&R$^2_{\Delta B}$ \vspace{0.1cm}\\
 &$\frac{R}{R_{\odot}}$&$\frac{R}{R_{\odot}}$&$\frac{R}{R_{\odot}}$&$\frac{R}{R_{\odot}}$&$\frac{R}{R_{\odot}}$\vspace{0.2cm}\\
\hline\noalign{\smallskip}
mod1 &34.7  &34.8  &34.3  & 37.5  & 36.3\\
mod2 &36.7  &36.6  &36.7  & 36.7  & 38.6\\
mod3 &59.1  &59.0  &58.6  & 59.5  &  60.9\\
mod4 &64.6  &65.7  &65.2  & 63.1  & 73.1\\
mod5 &82.1  &82.2  &81.0  & 81.3  & 78.0\\
mod6 &92.4  &92.3  &91.3  & 90.6  & 94.0\\
mod7 &98.6  &98.5  &97.3  & 99.4  & 99.6\\
mod8 &190.3 &188.8 &185.7 & 185.8 & 153.4\\
mod9 &206.9 &206.5 &204.1 & 202.0 & 203.1\\
mod10&223.1 &223.4 &220.6 & 187.7 & 177.0\\
\noalign{\smallskip}
\hline
\hline 
\end{tabular}
\end{table}

\begin{figure}
   \centering
   \includegraphics[clip,width=4cm,bb=19 171 300 700]{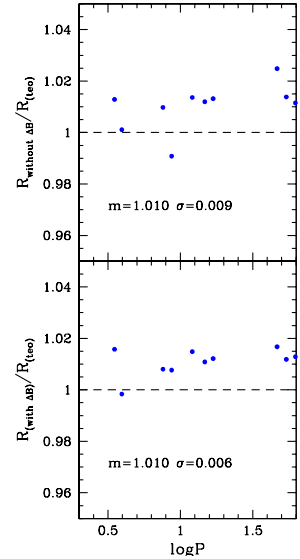}
      \caption{Ratio between ``computed'' and ``theoretical'' radii as a 
function of the logarithmic period. The bottom panel displays the radius 
evaluation based on the revised CORS method, while the top one 
the radius evaluations based on the pure BW method.  \label{DRteo1}}
         
   \end{figure}

The same result is also shown in Fig.~\ref{DRteo1}, which suggests  
that our calibration of the surface brightness 
is intrinsically correct with the discrepancy between  
``computed'' and ``theoretical'' Cepheid radii being around 1 $\%$. 
We also note that  the inclusion of the $\Delta B$ term only slightly improves the agreement with predicted radii, producing a small reduction of the scatter around the mean  (see labelled values in Fig.~\ref{DRteo1}).

\subsection{Stability tests}
Once verified that the new calibration of the surface brightness is 
intrinsically correct, we are in the position  to estimate the 
statistical error 
in the radius determination due to uncertainties in the measurements, i.e. 
to test the sensitivity of our technique to the accuracy of 
observed light, color and radial velocity curves. 
To this aim, we have transformed the synthetic curves for models in Table \ref{tabella1} by adopting the following steps: 
\begin{enumerate}
\item  we have reduced the number of 
phase points to the typical value for observations, i.e. $\sim 30-35$;
\item we have extracted phases randomly;
\item we have added Gaussian errors, considering the cases:\\
a) good data ($\sigma_V=0.02 mag$, $\sigma_{[m_1]}=0.04 mag$, 
$\sigma_{[c_1]}=0.04 mag$, $\sigma_{RV}=0.25 km^{-1}$, number of phase points=35);\\ 
b)  fair data ($\sigma_V=0.04 mag$, $\sigma_{[m_1]}=0.08 mag$, 
$\sigma_{[c_1]}=0.08 mag$, $\sigma_{RV}=0.5 km^{-1}$, number of points=30).
\end{enumerate}

\begin{figure}
   \centering
   \includegraphics[width=8cm]{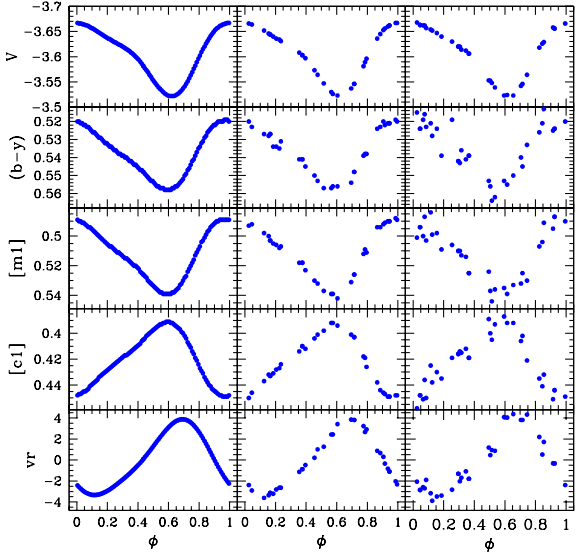}
      \caption{ Light, color and radial velocity curves for the model  (mod4) with 
$M=6.25 M_{\odot}$ and $T_{eff}=5100 K.$  The left panels show the pure model 
curves; the middle  and right ones the modified curves in the 
case of good and fair data respectively.  \label{datisfoltiti}}
         
   \end{figure}

As an example, Fig.~\ref{datisfoltiti} 
shows the synthetic light, color and radial velocity curves obtained 
for  model  mod4 (see Table \ref{tabella1}). Left panels show the pure model curves, 
whereas middle and right panels    
show the modified curves in the two cases  a) and b) respectively.

   \begin{figure}
   \centering
   \includegraphics[width=8cm]{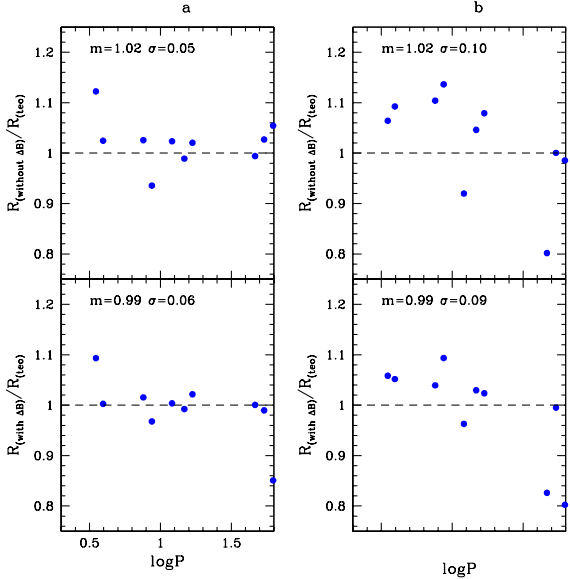}
      \caption{Ratio between ``computed'' and ``theoretical'' radii 
as a function of the logarithmic period. In the left panels (a) the radius is computed using modified curves that simulate good quality, adopting either the revised CORS method (bottom) or the pure BW one (top). In the right panels (b) similar plots are shown for radii computed from modified curves that simulate fair quality data.\label{DRteo2}}
         
   \end{figure}


We then applied the new CORS method to the whole set of modified 
synthetic curves. The resulting radii are reported in the last two columns of
Table \ref{tabella2}, whereas in Fig.~\ref{DRteo2} we plot  
the ratio between ``computed'' and ``theoretical'' radii as a function 
of the logarithmic period.  In particular, top and bottom panels 
show the CORS solutions
without and with $ \Delta B$ respectively for the cases a) (left) and 
b) (right).  
 Figure ~\ref{DRteo2} seems to show that including the $ \Delta B$ term does 
not improve much the results, on the contrary, the scatter in case a) worsen.
However, this disagreeable occurrence is only apparent. In fact, by excluding 
 ``mod8'' and ``mod10'' that show a very peculiar 
morphology (a sharp bump) of the $[c_1]$ color curve 
(see figure Fig.\ref{curveteo} for ``mod10''), which makes difficult the fit
and the $\Delta B$ calculation, the average uncertainties on the 
radius estimation fall to $3.5 \%$ and $5 \%$ 
in cases a)  and b) respectively (including the $\Delta B$ term).\\
These numbers represent a useful lower limit for the error 
associated with radius determination obtained by the CORS version developed 
in this paper. 

Another possible source of uncertainty in the radius determination 
is the misalignment between light or color curves, and the radial velocity one. 
In fact, the photometric and radial velocity data are rarely collected 
simultaneously. 
This occurrence could introduce  a shift in phase $\Delta \phi$  between the two different  data set. 
To verify the importance of this shift on the radius determination 
we have introduced an artificial phase shift 
(starting steps of  $\Delta \phi=0.01$  up to $\Delta \phi = 0.1$) 
in the synthetic radial velocity curves with respect to the photometric 
ones.

   \begin{figure}
   \centering
   \includegraphics[clip,width=6cm]{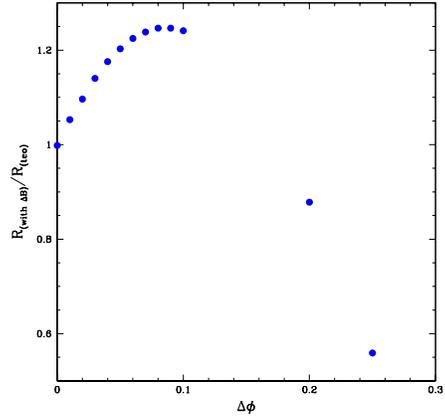}
      \caption{Ratio between the ``computed'' and ``theoretical'' 
radii as a function of the shift $\Delta \phi$ for the model 
(mod2) with $M=$ 5 $M_{\odot}$, $T_{eff}=5600 K$. \label{sfasamento}}
\end{figure}


The result of such a test is shown in  Fig.~\ref{sfasamento}, where the  
ratio between ``computed'' and ``theoretical'' radii is plotted 
as a function of the phase shift $\Delta \phi$ for the model mod2.
The ``computed'' radius is larger than the ``theoretical'' one 
for $0< \Delta \phi \leq 0.1$, whereas it becomes smaller beyond 0.1 with the effect increasing with the phase shift.  
Eventually, our program does not converge for $\Delta \phi \geq 0.3$. 
However, such large phase shifts are never reached when we deal 
with actual data. In fact,  when the temporal 
distance between photometric and radial velocity data is rather long  ($\sim$ 1000 cycles ), for a typical Cepheid  with period 
$P \sim 10 \pm 5 \cdot 10^{-5}$ d, the resulting phase shift 
is $\Delta \phi \sim 0.05$. As shown in Fig.~\ref{sfasamento}, this value 
of $\Delta \phi$ is sufficient to generate a systematic error 
on the derived radius of about $17 \%$. 
This occurrence confirms quantitatively the need to use photometric 
and radial velocity data as close as possible in time,  or otherwise to correct this shift.

\section{Application of the method}

Having tested the capabilities of the new version of the CORS method 
(see previous section), we are ready to 
 apply it to actual data. In the following we discuss the application to a sample of 52 Galactic Cepheids.

\subsection{The sample}

We searched the literature for an homogeneous sample of Galactic Cepheids with  photometric data in the Str\"omgren system and we selected the 31 pulsators
analysed by Arellano Ferro et al. (1998) (AFGR hereafter) supplemented by data
for other 21 objects from the papers by Feltz \& McNamara (1980) (FM hereafter) and Eggen (1985, 1996) (Eg  hereafter).

Concerning radial velocity data, several large and homogeneous datasets 
are available in the literature. In particular, 
we have used the catalogues from: Evans (1980) (E  hereafter), 
Gieren (1981) (G  hereafter), Barnes et al. (1987, 1988), (BMS  hereafter), 
Coulson et al. (1985) (CCG  hereafter), Wilson et al. (1989) (WCB  hereafter), 
Metzger et al. (1993) (MCMS  hereafter), Bersier et al. (1994) (BBMD  hereafter), Gorynya et al. (1998) (GSSRGA  hereafter), 
Kiss (1998) (K  hereafter) and 
Imbert (1999) (I  hereafter).\\

In general we considered only Cepheids with good photometric and radial 
velocity curves, 
i.e. more than 15 phase points and reasonable precision. 
When measurements by different authors are available for the same stars we have chosen data which have: 1) the largest phase points number and 
the best precision; 2) the lowest temporal separation with respect to the photometric 
data.  When we have distant data sets, that is differences between radial velocity and photometric curves larger than $\sim 1000$ cycles, (6 stars, see Table \ref{tabella3}) we correct the misalignment between this two curves by using different epochs.\\
 In few cases we have merged data from different authors,
 in order to obtain more sampled radial velocity curves 
(see Table \ref{tabella3}).
Our final selected dataset is summarized in Table~\ref{tabella3}.

\begin{table*}[t]
\begin{center}
\caption{The selected Cepheid sample  \label{tabella3}}
\begin{tabular}{c c c c c c c c c c }
\hline
\hline
Cepheid& Period & Photometry & Rad. Vel. & $\Delta (Ph-RV)$ & Cepheid & Period & Photometry & Rad. Vel.& $\Delta (Ph-RV)$\vspace{0.1cm} \\
 &days & source & source&cycles & &days & source & source&cycles\vspace{0.1cm}\\
\hline
FF Aql      &4.470840 &AFGR &BMS    &1142&YOph     &17.126780 &AFGR & GSSRGA&46\\
FM Aql      &6.114334 &AFGR &GSSRGA &89  &AW Per   & 6.463589 &FM   & GSSRGA&422\\
FN Aql      &9.481614 &AFGR &GSSRGA &56  &V440 Per & 7.572712 &Eg   & GSSRGA&131\\
$\eta$ Aql  &7.176779 &FM   &BMS    &413 &CM Sct   & 3.916977 &AFGR &M CMS, B&463\\ 
TT Aql      &13.75551 &AFGR &GSSRGA &81  &EV Sct   & 3.090998 &AFGR &BBMD, MCMS&586\\
U Aql       & 7.024100 &AFGR &BMS    &674 &RU Sct   & 19.70062 &AFGR & MCMS&92\\    
V496 Aql  &6.807164 & AFGR & G &643  &SS Sct   & 3.671280 &AFGR & BMS, G&1194\\
V600 Aql &7.238748 & AFGR & GSSRGA &156 &V367 Sct & 6.29307  &AFGR & MCMS&317\\
RT Aur   & 3.728561  & FM   &GSSRGA &216  &Y Sct    &10.341650 &AFGR &BMS&421\\
RX Aur   & 11.623537 & FM   &I &271    &BQ Ser   &4.316700  &AFGR &GSSRGA&926\\GI Car  & 4.431035  & Eg   &G &-     &S Sge    &8.382044  &FM   &GSSRGA&209\\
SU Cas  & 1.949322  & FM   &GSSRGA&-  &AP Sgr   &5.0574269 &AFGR &BMS, G&864 \\TU Cas  & 2.139298  & FM   &GSSRGA&-  &BB Sgr   &6.637115  &AFGR &GSSRGA&166\\ $\delta$ Cep& 5.366316  & FM   &BBMD &393&U Sgr    &6.745363  &AFGR &GSSRGA&81\\   
DT Cyg      & 2.499086  & AFGR &GSSRGA&2203  &V350 Sgr &5.154557  &AFGR &GSSRGA&157\\  
SU Cyg      & 3.845733  & AFGR &BMS&1432     &X Sgr    &7.012630  &AFGR &WCB&759\\  
X Cyg       & 16.386130 & AFGR &BBMD&244    &Y Sgr    &5.773400  &AFGR &BMS, WCB&885\\
VZ Cyg      & 4.864504  & FM   &BBMD&50    &YZ Sgr   &9.553606  &AFGR &BMS&419\\   
BG Cru      & 3.342503  & Eg   &E&943 &W Sgr    &7.595080  &AFGR &BBMD&542\\    
W Gem       & 7.913960  & FM   &I&257       &WZ Sgr   &21.849708 &AFGR &GSSRGA&50\\  
$\zeta$ Gem & 10.150780 & FM   &BBMD&179    &SZ Tau   &3.148727  &FM   &GSSRGA\&-\\   
BG Lac      & 5.331938  & FM   &I&114       &SV Vul   &44.999660 &AFGR &I&103\\
Y  Lac      & 4.323776  & FM   &I&46       &T Vul    &4.435532  &FM   &BBMD&530\\ 
Z  Lac      & 10.88554  & FM   &BMS,GSSRGA&273&U Vul  &7.990736  &AFGR &GSSRGA&142\\   
X Lac       & 5.444990  & FM   &BBMD &1.9   &X Vul    &6.319562  &FM   &GSSRGA&-\\  
T Mon       & 27.024649 & FM   &GSSRGA&-  &         &          &     & &\\
BF Oph      & 4.067695  & AFGR &BMS, G &1077  &         &          &     & &\\
\hline
\end{tabular}
\end{center}
Photometry sources: Arellano Ferro et al. (AFGR,1998); Feltz \& McNamara (FM,1980) and Eggen (Eg,1985,1996).\\
Radial velocity sources: Evans (E,1980); Gieren (G,1981); Barnes et al. (BMS,1987, 1988); Coulson et al. (CCG,1985); Wilson et al. (WCB,1989); 
Metzger et al. (MCMS,1993); Bersier et al. (BBMD,1994); Gorynya et al. (GSSRGA, 1998); Kiss (K,1998) and Imbert (I,1999).
 $\Delta (Ph-RV)$: number of pulsational cycles between photometric and radial velocity curves.
\end{table*}

\subsection{Peculiar stars}
Some Cepheids in our sample need to be 
discussed individually: 

\begin{itemize}
\item 
For the stars TT Aql, X Cyg, T Mon, RU Sct, Y Sct, WZ Sgr and SV Vul, 
we can calculate the radius only in the classical Baade-Wesselink approximation 
(without the $\Delta B$ term), because for these stars the loop in the 
$[m_1]$, $[c_1]$ plane does not  lie completely within 
the theoretical grids by C97a,b (see paragraph 3.1);
\item 
We have excluded from our sample the Cepheids GI Car, SU Cas, Y Lac,  
  SZ Tau, X Vul for the poor data quality.
\item 
Stars DT Cyg, FF Aql, BG Cru and V440 Per are first overtone pulsators 
(Antonello, Poretti \& Reduzzi, 1990).
\item 
Stars TU Cas, V367 Sct and BQ Ser are double mode Cepheids (Pardo \& Poretti, 1997 and references therein). 
\item 
Stars U Aql, FF Aql, V 496 Aql, $\eta$ Aql, RX Aur, SU Cas, $\delta$ Cep, GI Car, SU Cyg, VZ Cyg, $\zeta$ Gem, X Lac, Y Lac, Z Lac, BG Lac, T Mon, Y Oph, BF Oph, AW Per, Y Sct, S Sge, W Sgr, Y Sgr, WZ Sgr, AP Sgr, V 350 Sgr, SZ Tau, T Vul, U Vul and SV Vul are members of binary systems.  Note that this list is the result of a detailed analysis of Szabados's selection. (Szabados, 2003 and references therein).  
\end{itemize}


\subsection{Comparison with the literature}

\begin{figure}
   \centering
   \includegraphics[width=9cm]{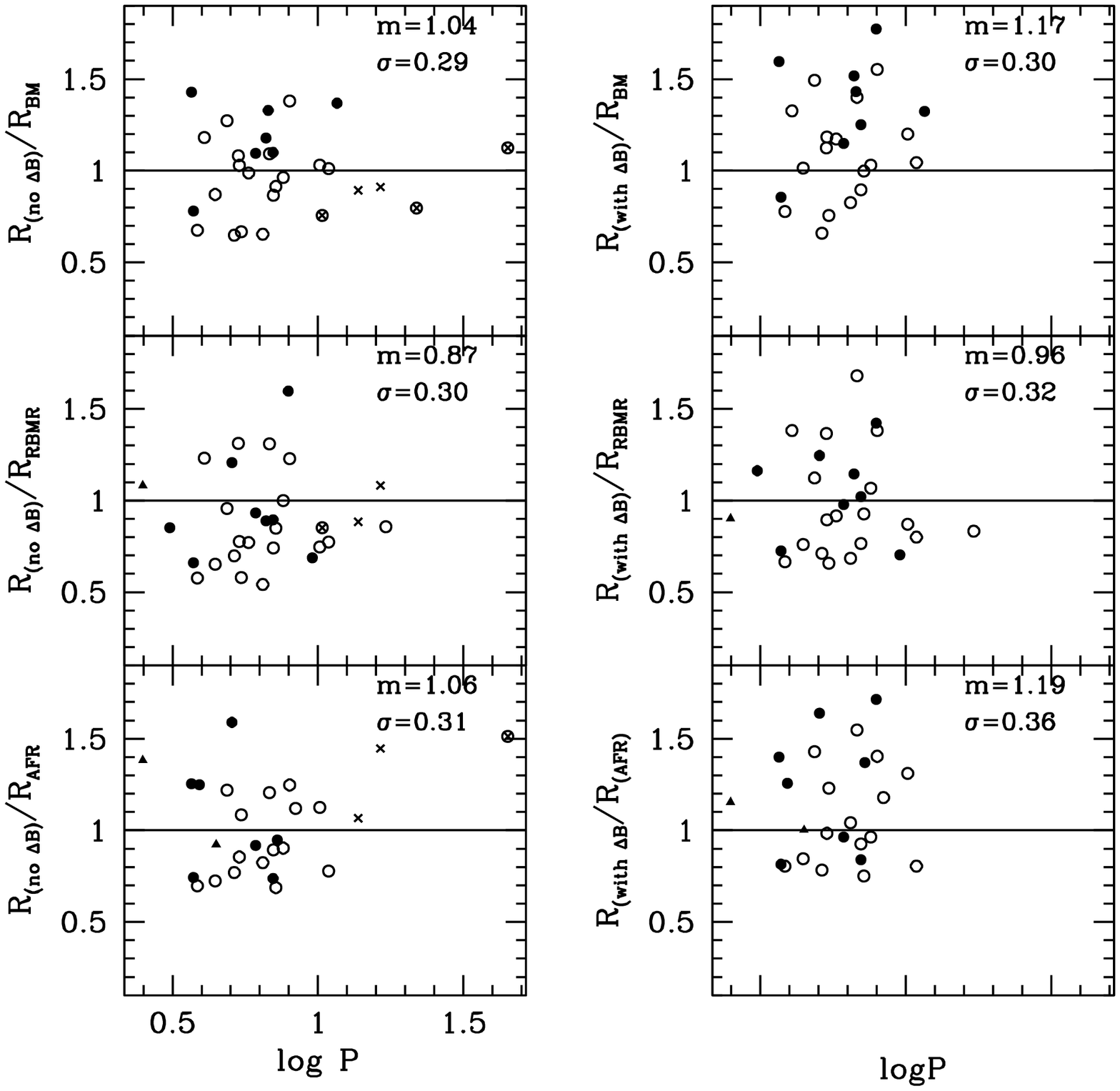}
      \caption{Comparison of radii obtained in this work ($R_{with \Delta B}$, $R_{no \Delta B}$) with those derived by other authors (Arellano Ferro \& Rosenzweing (2000) $R_{AFR}$;Ripepi et al. (1997) $R_{RBMR.}$;Moffett \& Barnes (1987) $R_{BM}$).  \label{confronto}}
         
   \end{figure}

In table~\ref{tabella4} we report the radius obtained for all the Cepheids in our sample, 
and, for comparison purposes, we report the literature results renormalized to 
our projection factor $k$. 
In particular, from left to right we report for each Cepheid: the  name; the period; 
the CORS radii obtained without and with the $\Delta B$ term 
(R$_{without \Delta B}$ and R$_{\Delta B}$); 
the radii obtained by Arellano Ferro \& Rosenzweing (2000) $R_{AFR}$, 
Laney \& Stobie (1995) $R_{LS}$ or  Gieren et al. (1998) $R_{GFG}$,
Ripepi et al. (1997) $R_{RBMR.}$ and Moffett \& Barnes (1987) $R_{BM}$.\\
In Fig.~\ref{confronto} we have compared our results with those of authors reported in 
the table \ref{tabella4}. As a result we find that our radii are on average 
larger than the radii obtained by Arellano Ferro \& Rosenzweing (2000) and  
Moffett \& Barnes (1987), while they are slightly smaller than Ripepi et al. (1997) ones 
(see labels in Fig.~\ref{confronto}).
The scatter in this comparisons is rather large ($\sim $ 30\%); this occurrence could 
be due to: 1) the inclusion of binary stars in the comparison (different methods 
uses different colours  and then binarity could affect  differentially 
the various determinations); 2) the use of optical colors. 
Verifying  hypothesis 1) by comparing only the non-binary stars 
does not make sense because of the small number of ``bona fide'' single stars
in our sample.

Concerning point 2), as  well known in the literature (Laney \& Stobie 1995, Gieren, Foqu\'e \& G\'omez 1997), radii 
obtained from NIR data are more precise than the ones obtained by using optical colors.
Unfortunately, most works with radii determinations from NIR data deal 
with southern Cepheids and we have very few stars in common (see column 7 of Table 4
for the comparison)

\begin{table*}[t]
\begin{center}
\caption{Radii derived with the current CORS method compared with 
previous determinations (see text) \label{tabella4}}. 
\begin{tabular}{c c c c c c c c c }
\hline
\hline
Cepheid &Binary & Period & R$_{without \Delta B}$ & R$_{\Delta B}$ & $R_{AFR}$ & $R_{LS}/R_{GFG}$ & $R_{RBMR.}$ & $R_{BM}$\vspace{0.1cm}\\
& &days &$\frac{R}{R_{\odot}}$ & $\frac{R}{R_{\odot}}$ & $\frac{R}{R_{\odot}}$ &$\frac{R}{R_{\odot}}$ & $\frac{R}{R_{\odot}}$ & $\frac{R}{R_{\odot}}$\vspace{0.1cm}\\
\hline
FF Aql$^{*}$ & O & 4.470840 &  49.8 &  54.1 & 52.0 & - & - & -\\
FM Aql       & - & 6.114334 & 56.5 & 59.3 & 59.3 & - & 60.6 & 51.61\\
FN Aql       &  - & 9.481614 & 83.5 & 86.5 & - & - & - & -\\
$\eta$ Aql   & B & 7.176779 & 48.2 & 52.6 & 67.5 & -& 56.7 & 52.76\\
TT Aql       & - & 13.75551 & 84.3 & - & 76.1 & - & 95.3 & 94.31\\
U Aql        &  O & 7.024100 & 45.4 & 47.0 & 48.9 & -& 61.3 & 52.45\\
V496 Aql     & B & 6.807164 & 47.7 & 61.2 & 38.1 & - & 36.4 & 43.69 \\
V600 Aql     & - & 7.238748 & 55.4 &8 0.1 & 56.3 & - & - & -\\
RX Aur       & B: & 11.623537& 83.7 & 81.0 & - & - & - & 61.17\\
RT Aur       & - & 3.728561 & 27.9 & 30.6 & 36.2& - & 42.2 & 35.74\\  
$\delta$ Cep & V & 5.366316 & 41.0 & 47.2 & 46.2 & - & 52.8 & 39.84\\
DT Cyg$^{*}$ & - & 2.499086 &  46.4 & 38.7 & 32.3 & - & 42.9 & -\\
SU Cyg       & O & 3.845733 &  35.4 & 40.8 & 48.9 & -& 61.3 & 52.45\\
X Cyg        & - & 16.386130&104.1 & - & 69.3 & - & 96.2 &114.29\\ 
VZ Cyg       & O & 4.864504 & 45.3 & 53.1 & 35.8 & -& 47.3 & 35.57\\
BG Cru$^{*}$ & - & 3.342503 & 28.9 & 32.7 & - & - & - & -\\ 
$\zeta$ Gem  & V & 10.150780 & 64.4 & 75.0 & 55.1 & - & 86.2 & 62.55\\
W Gem        & - & 7.913960 & 96.9 & 86.4 & 48.5 & - & 60.7 & 48.69\\
BG Lac       & b & 5.331938 & 47.0 & 48.9 & - & - & 35.8 & 43.49\\
X Lac        &  b & 5.444990 & 41.3 & 46.9 & 36.7 & - & 71.2 & 61.97\\
Z Lac        & O & 10.88554 & 67.2 & 69.4 & 83.1 & - & 86.8 & 66.40 \\ 
BF Oph       &  b & 4.067695 &  39.9 & 44.8 & - & 36.11$^a$/ 35.8$^b$ &32.4 & 33.78\\
Y Oph        &  b &17.126780 & 96.3 & 93.4 & - & 93.22$^a$ & 112.2 &-\\
AW Per       & O & 6.463589 & 29.7 & 37.5 & 34.7 & - & 54.8 & 45.40\\
V440 Per$^{*}$ & B: &7.572712 & 62.9 & 76.8 & - & - & - & -\\
CM Sct       & - & 3.916977 & 38.4 & 38.6 &29.6 & - & - & -\\
EV Sct       & B: & 3.090998 & 30.9 & 42.0 & -& 47.62$^a$/ 32.5$^b $ & 36.6 & -\\
SS Sct       & - & 3.671280 &  39.6 & 44.2 & 30.4 & - & - & 27.71\\ 
RU Sct       & - & 19.70062 &102.9 &- & - & 120.45$^a$  & - & -\\
Y Sct        & b &10.341650 & 60.9 & - & - & - & 71.5 & 80.45\\
AP Sgr       &  B: &5.0574269 & 60.4& 62.3 & 36.6 & - & 50.0 & - \\
S Sge        & O & 8.382044 & 75.1 & 79.1 & 64.6 & - & - & -\\
BB Sgr       & - & 6.637115 & 47.6 & 61.3 & - & 37.39$^a$ /44.4$^b$ & 53.5 & 40.42\\
U Sgr        &  - & 6.745363 & 76.6 & 82.5 & - & 52.30$^a$/ 48.8$^b$ & - & 57.61\\
V350 Sgr     & O & 5.154557 & 32.1 & 32.7 & 40.2 & - & 46.0 & 49.55\\
X Sgr        & - & 7.012630 & 52.5 & 59.8 & 68.5 & - & 58.6 & 47.76\\
Y Sgr        & B & 5.773400 & 47.4 & 56.2 & - & - & 61.4 & 47.95\\
YZ Sgr       &  - & 9.553606 & 93.6 & 95.8 & - & - & 136.0 & -\\
W Sgr        &  O & 7.595080 & 58.6 & 62.6 & 62.5 & - & 58.6 & 60.77\\
WZ Sgr       & B &21.849708 &120.5 & -    & - & 121.43$^a$ /122.2$^b$ & - & 151.20\\
SV Vul       & B &44.999660 &213.0 & - & 135.6 & 241.62$^a$/ 250.7$^b$ & - &189.5\\
T Vul        & b &4.435532 & 31.8 & 37.1 & 42.3 & - & 48.8 & 36.54\\
U Vul        & O &7.990736 & 75.0 & 84.4 & 57.9 & - & 61.1 & 54.34\\
\hline
\end{tabular}
\end{center}
An asterisk at top-right of Cepheid name means that it pulsates as a 
First Overtone. \\
In the second column: B - spectroscopic binary, B: - spectroscopic binary but 
confirmation needed; b - Photometric companion, 
physical relation should be investigated; O - spectroscopic binary 
with known orbit; V visual binary (see Szabados, 2003 for details).\\
Radius sources: $R_{AFR}$ Arellano Ferro \& Rosenzweing (2000); $R_{LS}$ Laney \& Stobie (1995) / $R_{GFG}$ Gieren et al. (1998) in the table $^a$ refer to $R_{LS}$ and  $^b$ to $R_{GFG}$; $R_{RBMR}$ Ripepi et al. (1997); $R_{BM}$ Barnes \& Moffet (1987).
\end{table*}

Recently it has become possible to derive accurate stellar radii from the 
angular diameters measured with interferometric  
 techniques (see Nordgren et al. 2000 and Lane et al. 2002) combined with  
\textit {Hypparcos} parallaxes. The comparison with these measurements represents a useful 
test for our results. 
In Table \ref{confrNord} we compare with our results the radii obtained by Nordgren al. (2000)
and  Lane et al. (2002), with the quoted interferometric techniques, for the stars in 
common with our sample, namely $\eta$ Aql, $\delta$ Cep and $\zeta$ Gem. In particular, 
from left to right, we report:  the name of the star;
the radius  ($R_I$) obtained by means of interferometric techniques ($\delta$ Cep 
from Nordgren et al. 2000 and $\eta $Aql and $\zeta$ Gem from 
Lane et al. 2002); our radius in the two different approximations, 
without ($R_{without \Delta B}$) and with ($R_{with \Delta B}$) the $\Delta B$ term respectively. The interferometric radii, reported in Table \ref{confrNord}, are corrected for the different $k$ projection factors. So we obtain the radii 61.8 x 1.36/1.43= 58.8 $R_{\odot}$, 45x1.36/1.31=47 $R_{\odot}$ and 66.7 x 1.36/1.43=63.4 $R_{\odot}$ for $\eta$ Aql, $\delta$ Cep and $\zeta$ Gem, respectively. 
 \begin{table}[t]
\begin{center}
\caption{\footnotesize{Comparison with radii measured with interferometric techniques (see text for details).   \label{confrNord}}}
\small
\begin{tabular}{c c c c}
\hline
\hline
Star&$R_I$&$R_{without \Delta B}$&$R_{with \Delta B}$ \vspace{0.1cm}\\ 
\hline
$\eta$ Aql   &$58.8 \pm 7.6$  &48.2 &52.6\vspace{0.1cm}\\ 
$\delta$ Cep &$47^{+8}_{-6}$  &41.0 &47.2\vspace{0.1cm}\\
$\zeta$ Gem  &$63.4 \pm 7.2$ &64.4 &75.0\vspace{0.1cm}\\
\hline
\end{tabular}
\end{center}
\end{table}
We notice that the agreement between our results and the interferometric ones is very 
good for $\eta$ Aql and $\delta$ Cep, whereas some discrepancy is found for $\zeta$ Gem . However, by 
assuming a typical error of  $\sim 10 \%$ (see discussion in Sect.7) on the radii found, we conclude that our 
results are globally consistent with the interferometric ones.

\section{The Period-Radius relation}

After having tested the method and the consistency of our results with 
previous determinations in the literature, we are in the position 
to derive a PR relation based on the new derived radii.

\begin{figure}
   \centering
   \includegraphics[width=9cm]{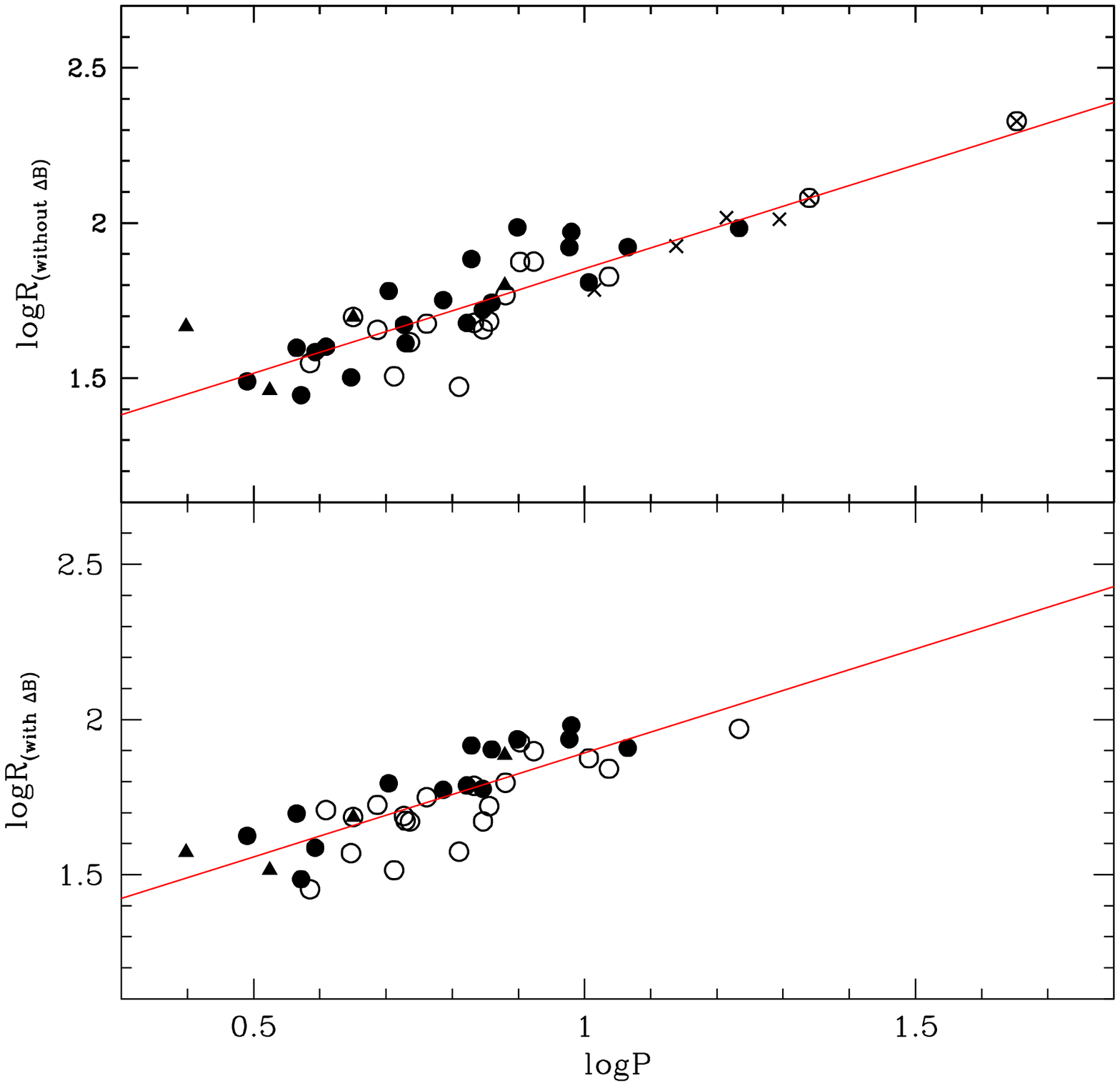}
      \caption{Top: Period-Radius relation obtained in the case without 
$\Delta B$; Bottom: as before, but with $\Delta B$. 
Open circle: Binary stars; Triangles: first overtone pulsators; 
Crosses: stars with loop in the $[m_1]$, $[c_1]$ plane not 
completely within the theoretical grids. see par.3\label{PRtutte}}
         
   \end{figure}

The PR relations for all the stars in our sample for which the 
program reached the convergence are shown in Fig.~\ref{PRtutte}, where 
the the top and bottom panels illustrate the case without and 
with $\Delta B$ respectively. 
A least square fit to the data (solid lines in Fig.~\ref{PRtutte}) 
 leads to the following Period-Radius relations in the case 
without the $\Delta B$ term (Eq.~\ref{eq:PRmiatuttesDB}) and 
with the $\Delta B$ term (Eq.~\ref{eq:PRmiatuttecDB}):

\begin{equation}
\label{eq:PRmiatuttesDB}
 \log R = (1.18 \pm 0.05) + (0.67 \pm 0.06) \log P \\\\ r.m.s=0.09  
\end{equation}

\begin{equation}
\label{eq:PRmiatuttecDB}
 \log R = (1.22 \pm 0.06) + (0.67\pm 0.08) \log P \\\\ r.m.s=0.08
\end{equation}

\begin{figure}
   \centering
   \includegraphics[width=9cm]{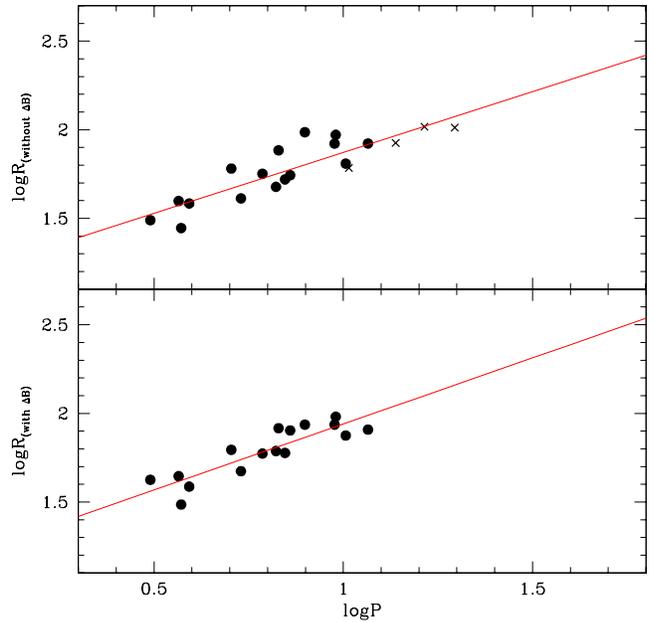}
      \caption{Top: The same of figure \ref{PRtutte} but excluding  first overtone and binary stars. \label{PRarticolo}}
         
   \end{figure}

An inspection of the figure suggests that the presence 
of first overtone pulsators (filled triangles) could affect the 
derived PR relations because, at fixed period, they are brighter  
and, in turn, have larger radii than fundamental pulsators. \\
Similarly, also Cepheids belonging to binary systems could 
affect the radius determination.  Therefore, we decided to exclude 
from our relations the first overtone pulsators and  the stars flagged as ``B'' and ``O''\footnote{We tried to determine the radius for the few 
Cepheids with known orbits, by disentangling orbital motions and 
pulsation in the radial velocity curve, but the error produced 
in this procedure remained still too high to secure 
good results.} in Table~\ref{tabella4},  whereas we left in our 
sample the Cepheids flagged as ``B:'', ``b'' and ``V'' (i.e. uncertain  
spectroscopic binaries and separated visual binaries respectively, 
see Szabados 2003). 
 As a result of this selection procedure, 
we are left with 20 and 16 Cepheids in the cases without and with 
$\Delta B$ respectively. We therefore calculated new Period-Radius 
relations with the following
results (see Fig.~\ref{PRarticolo}): 

\begin{equation}
\label{eq:PRmiasDB}
 \log R = (1.18 \pm 0.08) + (0.69 \pm 0.09) \log P  \\\\   r.m.s=0.08
\end{equation}    

\begin{equation}
\label{eq:PRmiacDB}
 \log R = (1.19 \pm 0.09) + (0.74 \pm 0.11) \log P   \\\\   r.m.s=0.07
\end{equation}

A comparison between Eq.~\ref{eq:PRmiatuttesDB},~\ref{eq:PRmiatuttecDB} and 
Eq.~\ref{eq:PRmiasDB},~\ref{eq:PRmiacDB} shows that the net result of our 
selection criterion  mainly consist  in increasing the slope of 
the PR relation, going in the direction to improve the agreement with  the literature, and, in particular with the NIR results by 
 Laney \& Stobie (1995) and Gieren et al. (1998) 
(see table \ref{PRlett} for a comparison between present results and other relations in the literature). Yet, we have to note that the errors 
of the derived coefficients are rather large. This is caused by: 
1) the small number of Cepheids left after the selection process (In particular the number of binary stars in our sample is very large)
2) the lack of long period Cepheids in the case with $\Delta B$, as 
a result of problems with synthetic model atmosphere grids (see Sect. 3.1).
We remark, however,  that including or not the $\Delta B$ term 
in the CORS determination makes some difference, in the sense that, apart from
a few Cepheids in the low period range, the scatter in the PR relation 
is slightly reduced (see Fig.~\ref{PRarticolo}) when the $\Delta B$ term is included. This result  seems to suggest that  the modification to the CORS method presented in this 
paper is well suited also when applied to real stars (not merely to synthetic 
light curves).

\begin{table*}[t]
\begin{center}
\caption{{Comparison between the PR ($\log R = a \log P +b$) coefficients 
obtained in this paper and the ones based on selected works in literature. 
In particular, from left to right, we report the slope, 
the zero point, the source and the method adopted for deriving the PR.
\label{PRlett}}} 
\begin{tabular}{l l l l l}
\hline
\hline
a&b&Source&Method\vspace{0.1cm}\\ 
\hline         
0.74 $\pm$ 0.03&1.12 $\pm$ 0.03&RAAF   &Surf. Brightness\vspace{0.1cm}\\
0.751 $\pm$ 0.026&1.070 $\pm$ 0.008&LS &Surf. Brightness\vspace{0.1cm}\\
0.750 $\pm$ 0.024&1.075 $\pm$ 0.007&GFG&Surf. Brightness\vspace{0.1cm}\\
0.606 $\pm$ 0.037&1.263 $\pm$ 0.033&RBMR&CORS \vspace{0.1cm}\\
0.655 $\pm$ 0.006&1.188 $\pm$ 0.008&BCM&Theory\vspace{0.1cm}\\ 
 0.69 $\pm$ 0.09& 1.18 $\pm$ 0.08&This work & new CORS (without $\Delta B$)\vspace{0.1cm}\\
 0.74 $\pm$ 0.11& 1.19 $\pm$ 0.09&This Work & new CORS (with $\Delta B$)\vspace{0.1cm}\\
\hline
\end{tabular}
\end{center}
PR source: Rojo Arellano \& Arellano Ferro (1994, RAAF); Laney \& Stobie (1995, LS); 
Gieren et al. (1998, GFG); Ripepi et al. (1997, RBMR); 
Arellano Ferro \& Rosenzweig (2000, AFR);Bono et al. (1998, BCM). 
\end{table*}

Finally, we notice that our results concerning the radii and the PR relations 
could be, in principle, combined with an effective temperature 
calibration to derive  the intrinsic stellar luminosity, 
through the Stefan-Boltzmann law,  and in turn distance to studied 
Cepheids. This possibility will be investigated in a forthcoming paper.
In the following section we will apply an alternative method to derive the  
luminosity and distance by means of the comparison of empirical light and 
radial velocity curves with the predictions of the nonlinear convective 
pulsation models discussed above.

\subsection{Theoretical Fit of the light and radial velocity 
curves of the Cepheid Y Oph}

Before going to the conclusions of this paper, we note that 
it has recently been suggested that nonlinear pulsation models 
provide a direct tool to evaluate the intrinsic stellar properties 
of pulsating stars through the comparison of observed and predicted 
variations of relevant parameters along a pulsation cycle (see Wood, 
Arnold \& Sebo 1997, Bono, Castellani \& Marconi 2000, 2002).
This kind of analysis also provides an additional test for our radius 
determination technique by means of the comparison with the radius of 
the model which is able to simultaneously reproduce the period, the 
amplitude and the morphology of light and radial velocity curves.
We plan to apply this method to a sample of Galactic Cepheids with 
accurate photometric data  and available radial velocity information. 
In this paper we present a first application to the Cepheid Y Oph.
The observed properties of this star are summarized in 
Table~\ref{tabunica}.\\

\begin{table}[htpb!]
\begin{center}
\footnotesize
\begin{tabular}{c c c c c c}
\hline
\hline
P&$<m_V>$&$B-V$&Z   &E(B-V)&A\\
\hline
17.1268&6.169&1.377&0.05&0.655&0.483\vspace{0.1cm}\\
\hline
\hline
M&Y&Z&$T_{eff}$&$\frac{R}{R_{\odot}}$&$\log \frac{L}{L_{\odot}}$\vspace{0.1cm}\\
\hline
7 $M_{\odot}$&0.28&0.02&4720 K& 97.77&3.64\vspace{0.1cm}\\
\hline
\end{tabular}
\caption{\footnotesize {Top: Observed properties of Y Oph; Bottom: Physical parameters of the best fit model\label{tabunica} }} 
\end{center}
\end{table}

\begin{figure}[t]
\begin{center}
\includegraphics[clip,width=8cm]{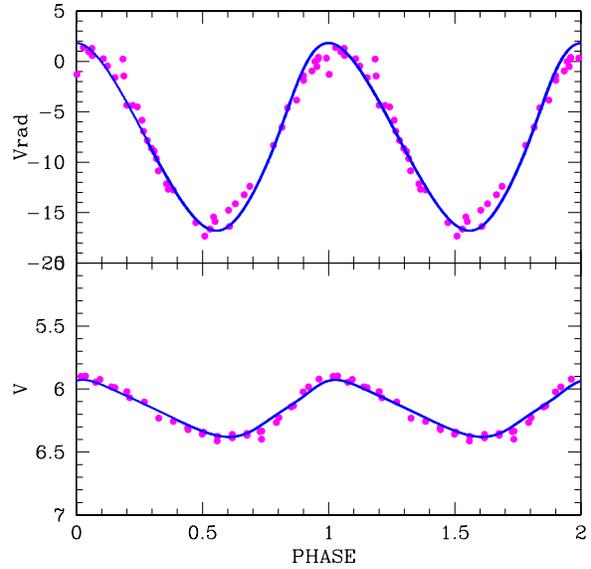}
\caption{\footnotesize{Top panel: empirical light curve (dots) 
for the star Y Oph, compared with the best fit model (solid 
lines, see text for details). Bottom panel: the same but 
for the radial velocity curve. \label{fitmod}}} 
\end{center}
\end{figure}

Starting from the observed radial velocity and light curve, 
we try to reproduce their morphology and amplitude, 
by computing pulsation models along isoperiodic (with period 
equal to the observed one) sequences with varying pulsation 
mass and effective temperature and assuming, for each mass, 
a canonical mass-luminosity relation. 
The best fit model resulting from these computations is shown 
in Fig. \ref{fitmod} (solid line). The corresponding stellar 
parameters are reported in Table \ref{tabunica} and show that 
the agreement with the radius determined in this paper is good.

The other important information provided by the fit is the 
stellar absolute magnitude $M_V=-3.996$ mag,   which  allows 
us to estimate a distance of about $423$ pc, consistent with 
the independent evaluation by Gieren, Barnes \& Moffett (1993), thus 
supporting the predictive capability of pulsation models.

\section{Final remarks}

We have presented a modified version of the CORS 
method based on a new calibration of the Surface Brightness 
function in the Str\"omgren photometric system. 
In particular we have been able to derive a calibration of 
$S_V$ as a function of the Str\"omgren reddening free 
indexes $[m_1]$ and $[c_1]$ by adopting grids of 
theoretical colors. This procedure revealed the unexpected occurrence 
that the quoted theoretical grids are not able to fully 
include the location of actual long period (P $\ge$ 12-13 days) 
Cepheids loops in the $[m_1]$,$[c_1]$  plane. This 
problem could only be overcome by adopting next generation, 
hopefully improved, model atmosphere grids.
Nevertheless, the modified CORS method presented here 
has been tested by means of synthetic light and radial velocity 
curves derived from nonlinear pulsation models and simulations have been 
performed to take into account the quality of real observed curves 
as well as possible time shifts between photometric and radial velocity data.
The results of such tests can be summarized as follows:\\
a) the present method appears capable 
to derive the radius of a Cepheids with an 
average precision of around  3.5\% and 5\% in the case of good and fair 
observational data respectively  (see Section 4.3);\\
b) the error associated with  a phase shift between photometric and 
radial velocity data could be as large as $\sim$ 17\% in the extreme 
case of  $\Delta \phi \sim 0.05$. In a  more common case of 
$\Delta \phi \sim 0.01$ the relative error is still rather large, namely 
$\sim$ 5\%. If we add to the error budget the uncertainty on the 
value of the projection factor $k$ 
($\sim \pm$ 2\% for $\Delta k \pm 0.03$), neglecting other 
possible contributions (see Bono, Caputo \& Stellingwerf 1994 and references therein for a detailed discussion), we
 can conclude that estimating  
the radius of a Cepheid by means of the Baade-Wesselink method 
(at least in the present formulation) with an accuracy better than 
10\% is extremely difficult to be achieved.\\

The method has been then applied to a sample of Galactic Cepheids 
with Str\"omgren photometry and radial velocity data to derive the 
radii and a new PR relation. As a result we obtained the following 
Period-Radius relation:
 $\log R = (1.19 \pm 0.09) + (0.74 \pm 0.11) \log P$ (r.m.s=0.07). 
This relation, even if not very accurate (mainly due to 
the unexpected presence of a large fraction of binary stars in our 
Cepheid sample), nevertheless is in satisfactory agreement 
with previous findings in the literature.

\begin{acknowledgements}
We wish to thank our anonymous referee for several pertinent suggestions 
that improved the content and the readability of the paper. 
This work made use of the ``McMaster Cepheid Photometry 
and Radial Velocity Data Archive'' maintained by Doug Welch, and of 
SIMBAD database, maintained at the CDS-Strasbourg.  This work was partially supported by MIUR/Cofin 2002, under the 
project ``Stellar Populations in Local Group Galaxies'' (Monica Tosi
coordinator).    
\end{acknowledgements}

\appendix
\section{}
The $4^{th}$ degree polynomial fit to effective temperature, effective gravity, and bolometric corrections mentioned in Sec.3.1.1 are the following:
\begin{eqnarray}
\log T_{eff} &=& a_0 + a_1 \cdot [m_1] + a_2 \cdot [m_1]^2+ a_3 \cdot [c_1] \\
\nonumber
&+&  a_4 \cdot [m_1] \cdot [c_1] + a_5 \cdot [m_1]^2 \cdot [c_1] + a_6 \cdot [c_1]^2 \\
\nonumber
&+&  a_7 \cdot [m_1]^2 \cdot [c_1]^2
\nonumber
\end{eqnarray}

\begin{eqnarray}
\log g_{eff} &=& b_0 + b_1 \cdot [m_1] + b_2 \cdot [m_1]^2+ b_3 \cdot [c_1] \\
\nonumber
&+&  b_4 \cdot [m_1] \cdot [c_1] + b_5 \cdot [c_1] ^2 + b_6 \cdot [m_1]^2 \cdot[c_1]^2 \\
\nonumber
\end{eqnarray}

\begin{eqnarray}
BC &=& c_0 \cdot \log \ T_{eff} + c_1 \cdot \log \ T_{eff}^2\\
\nonumber
&+& c_2 \cdot \log \ g_{eff} + c_3 \cdot \log \ T_{eff} \cdot\log \ g_{eff}\\
\nonumber
&+& c_4 \cdot \log \ T_{eff}^2 \cdot\log \ g_{eff}+ c_5 \cdot \log \ g_{eff}^2\\
\nonumber
&+& c_6 \cdot\log \ T_{eff}\cdot \log \ g_{eff}^2 + c_7 \cdot \log \ T_{eff}^2\\
\nonumber
&\cdot& \log \ g_{eff}^2
\end{eqnarray}
the coefficients $a_i$, $b_i$, $c_i$ of the previous relations are listened in Table A.1 and the other symbols have their usual meaning. Note the r.m.s of the previous relations are 0.0018 dex. , 0.1 dex. and 0.003 mag. respectively.
 
\begin{table}[b]
\begin{center}
\caption{Coefficients for the polynomial fits described in the Appendix.}
\begin{tabular}{rrrrrrrrr}
\hline
\hline\noalign{\smallskip}
$a_0$ &$a_1$  &$a_2$ &$a_3$ &$a_4$  &$a_5$ &$a_6$  &$a_7$ &$a_8$ \\
3.8911&-0.4273&0.1806&0.2168&-0.7049&0.5227&-0.0609&0.3989&  \\
\hline\noalign{\smallskip}
$b_0$ &$b_1$&$b_2$&$b_3$&$b_4$&$b_5$&$b_6$&$b_7$&$b_8$ \\
2.4   &11.7 &-7.1 &10.4 &-47.1&-3.5 &39.9 &     & \\
\hline\noalign{\smallskip}
$c_0$   &$c_1$  &$c_2$  &$c_3$ &$c_4$  &$c_5$&$c_6$  &$c_7$&$c_8$  \\
-276.874&144.304&-18.803&76.782&-40.744&5.403&-15.724&8.378&-1.116\\
\hline
\hline\noalign{\smallskip}

\end{tabular}
\end{center}
\end{table}

\end{document}